\documentclass[prb,showpacs,twocolumn]{revtex4}
\usepackage{graphicx}
\usepackage{dcolumn}
\usepackage{bm}
\usepackage{amsmath}
\usepackage{txfonts}

\begin{document}

\title{Spin-orbit interaction effect on transport of Dirac fermions in graphene}
\author {Kai-He Ding$^1$}
\email{dingkaih@mails.gucas.ac.cn}
\author{Guanghui Zhou$^2$}
\email{ghzhou@hunnu.edu.cn}
\author {Zhen-Gang Zhu$^3$}

\affiliation{$^1$Department of Physics and Electronic Science,
Changsha University of Science and Technology, Changsha, 410076,
China}

\affiliation{$^2$Department of Physics, Hunan Normal University,
410081, China and\\ International Center for Materials Physics,
Chinese Academy of Sciences, Shenyang 110015, China}

\affiliation{$^3$Institut f\"{u}r Physik
Martin-Luther-Universit\"{a}t Halle-Wittenberg
Nanotechnikum-Weinberg, Heinrich-Damerow-Strasse 4 D - 06120 Halle
(Saale), Germany}

\begin{abstract}
We study theoretically the quantum transport properties of the Dirac
fermions with spin-orbit interactions (SOIs) in graphene by using
the method of Schwinger proper time together with decomposition over
Landau level poles and Kubo formula. The analytical expressions for
both longitudinal and Hall conductivities are derived explicitly. It
is found that, from some numerical examples, when the Rashba SOI is
taken into account the Shubnikov-de Haas (SdH) oscillation peaks of
the longitudinal conductivity versus the chemical potential are
split, while the SdH oscillation of the longitudinal conductivity
versus a external magnetic field exhibits a beating pattern.
Furthermore, the Rashba SOI tends to suppress the quantum Hall
effect in graphene.
\end{abstract}
\maketitle
\section{Introduction}
Graphene has attracted a lot of attention because of its appealing
properties\cite{novoselov2}\cite{zhang}
\cite{geim}\cite{heersche}\cite{berger}\cite{castrocond}. At low
energy, owing to the specific band structure with the unique valley
and neutrality separating the hole states from the electron
states\cite{mcclure}, graphene has led to the emergence of a
paradigm of the relativistic condensed matter
physics\cite{gordon}\cite{haldane}, where the relativistic quantum
phenomena, some of which are unobservable in high energy physics,
can be tested in the table-top experiment. The recent advances in
fabrication technique have made it possible to produce graphite
systems with a few layers or even a single monolayer of
graphene\cite{novoselov1}\cite{Novoselov2005pnas}\cite{zhang2005prl}\cite{berger2004jpcb},
which propel graphene study into thriving high tide.

The relativistic feature of the graphene represents the anomalous
physical phenomena, such as anomalous quantized Hall effect,
absence of the weak localization and existence of the minimal
conductivity\cite{geim}. In addition to dissipative transport also
supercurrent transport has already been observed\cite{heersche}.
It is suggested that the graphene is a promising candidate for the
spintronics and related applications due to the
SOIs\cite{trauzettel}\cite{huertas}. In a recent paper, Kane and
Mele have studied the effect of the SOIs in
graphene\cite{kane1}\cite{kane2}, and find that the spin Hall
conductivity is quantized in the absence of a magnetic field
because of a gap produced by the SOIs. Depending on the relative
strength of intrinsic and Rashba SOIs, it is further shown that
the spin Hall conductivity can be zero or nonzero. Kane and Mele
have roughly estimated the SOI scale. Following this routine, Yao
{\it et al.}\cite{yao} and Min {\it et al.}\cite{min} have found
that this estimation is too big, and they have given some explicit
expressions of the SOIs in graphene.

In this work, we investigate the transport of Dirac fermions in
graphene. The main purpose is focused on the effect of the SOIs on
the transport on the basis of estimation of Yao {\it et al.} and
Min {\it et al.} for the SOIs in graphene. Using the Schwinger
proper-time method\cite{schwinger}, decomposition over Landau
level poles\cite{chodos}\cite{gusy} and Kubo
formula\cite{mahan.book}, we obtain some analytical expressions
for both longitudinal and Hall conductivities. It is found that
when the Rashba SOI is considered, the longitudinal conductivity
as a function of the chemical potential deviates the linear
relation at zero magnetic field. For nonzero magnetic field, the
SdH oscillations are observed and the oscillation peaks in the
longitudinal conductivity versus the chemical potential are split
when the Rashba SOI is applied, while the oscillation in the
longitudinal conductivity versus the magnetic field exhibits a
beating pattern. It is also shown that the Rashba SOI tends to
suppress the quantum Hall effect in graphene.

The rest of this paper is outlined as follows: In Sec. II, The
model of a single layer of graphite (graphene) with the SOIs is
established. In Sec. III, we derive the analytical expressions for
both the longitudinal conductivity and the Hall conductivity
including the limits of these expressions at zero field. In Sec.
IV, the corresponding numerical results and discussions are given.
In Sec. V, a summery is presented. Finally, Some tedious algebra
is included in the Appendix.

\section{Model Formalism}
The graphene is a flat monolayer of carbon atoms tightly packed into
a honeymoon lattice. At low energy, it can be described by 2+1
dimensional relativistic field theory model. When the SOIs are
included, the  Lagrangian density of the system is given by
\begin{equation}
\mathcal{L}=\hbar
v_F\overline{\Psi}(i\hat{D}+H_{s})\Psi,\label{lag}
\end{equation}
where $\Psi=(\Psi_K,\Psi_{k'})$ is the eight-component Dirac spinors
with $\Psi_{K(K')}=\left(\Psi_{A\uparrow},\Psi_{A\downarrow},
\Psi_{B\uparrow},\Psi_{B\downarrow}\right)$ which describes the
spin-related Bloch states residing on the atoms of the A, B
sublattice at momentum $K(K')$,
$\hat{D}=\gamma^\mu(\partial_\mu-ieA_\mu)$ with $\gamma^\mu$
($\mu=0,1,2$) being $4\times4$ $\gamma$ matrices\cite{gusynin2}, $e$
is the electron charge, $v_F$ is the Fermi velocity, the external
magnetic field $\mathbf{B}=\nabla\times \mathbf{A}$ is applied
perpendicular to the $x-y$ plane and the corresponding vector
potential is taken in the symmetric gauge $\mathbf{A}=(-By/2,Bx/2)$.
In Eq. (\ref{lag}), $H_{s}$ describes the SOIs that read
\cite{yao}\cite{min}
\begin{equation}
H_{s}=\lambda_{SO}(1-\gamma^0 s_z)+\lambda_R(i\gamma^1
s_y-\gamma^0\gamma^1 s_x),
\end{equation}
where $\lambda_{SO}$ is the intrinsic SOI parameter, $\lambda_R$
is the Rashba SOI parameter, and $\mathbf{s}$ is the spin
variable. For $B=0$, the corresponding energy spectrum are given
by
\begin{eqnarray}
\varepsilon_{1}=\pm\sqrt{\mathbf{k}^2+(\lambda_R-\lambda_{SO})^2}+\lambda_R-\lambda_{SO},\nonumber\\
\varepsilon_{2}=\pm\sqrt{\mathbf{k}^2+(\lambda_R+\lambda_{SO})^2}-\lambda_R-\lambda_{SO}.
\end{eqnarray}
For $\lambda_{SO}>\lambda_R>0$, the system includes an energy gap of
$2(\lambda_{SO}-\lambda_R)$. For $0<\lambda_{SO}<\lambda_R$, the
energy gap closes.

The Green's function of Dirac fermions described by the Lagrangian
(\ref{lag}) in an external magnetic field can be expressed
as\cite{schwinger}\cite{chodos}\cite{gusy}
\begin{equation}
G(x,y)=(i\hat{D}-H_{s})_x\langle
x|\frac{-i}{H_{s}^2+\hat{D}^2-i[\hat{D},H_{s}]}|y\rangle.
\end{equation}
Using the Schwinger proper time approach\cite{schwinger}, we
obtain
\begin{equation}
G(x,y)=\exp(ie\int_y^x A_\lambda
dz^\lambda)\widetilde{G}(x-y),\label{gxy}
\end{equation}
\begin{widetext}
\begin{equation}
\widetilde{G}(x)=\int_0^\infty ds\frac{e^{-\frac{i\pi}{4}}}{8(\pi
s)^{3/2}}e^{-\frac{i}{4s}x_\nu
C^{\nu\mu}x_\mu}[\frac{1}{2s}\gamma^\mu
C_{\mu\nu}x^\nu-\frac{1}{2}(e\gamma^1Bx_2-e\gamma^2
Bx_1)+\lambda_{SO}(1-\gamma^0s_z)]\frac{eBs}{\sin(eBs)}e^{i(\frac{1}{2}e\sigma
F-\Delta-B_\mu^2)s},
\end{equation}
\end{widetext}
where $F_{\mu\nu}=\partial_\mu A_\nu-\partial_\nu A_\mu$,
$C^{\mu\nu}=g^{\mu\nu}+\frac{(F^2)^{\mu\nu}}
{B^2}[1-eBs\cot(eBs)]$ with $g^{\mu\nu}=\text{diag}(1,-1,-1)$, and
\begin{equation}
B_\mu=i2\lambda_{SO}\sigma^{\mu 0}s_z+2\lambda_R(\sigma^{\mu
1}s_y-\delta_{\mu0}\gamma^1s_x-\delta_{\mu1}\gamma^0s_x),
\end{equation}
\begin{equation}
\Delta=2(\lambda_{SO}^2+\lambda_R^2)(1-\gamma^0
s_z)+4\lambda_{SO}\lambda_R(i\gamma^1 s_y-\gamma^0\gamma^1 s_x)
\end{equation}
with
$\sigma^{\mu\nu}=\frac{i}{2}(\gamma^\mu\gamma^\nu-\gamma^\nu\gamma^\mu)$.
It is clear that the symmetric gauge sets the factor
$\exp(ie\int_y^x A_\lambda dz^\lambda)$=1 in Eq. (\ref{gxy}).
Therefore, Eq. (\ref{gxy}) becomes the functions of the difference
$x-y$ only. Using the expansions of the exponential function $e^x$,
one can show that
\begin{widetext}
\begin{eqnarray}
e^{i(\frac{1}{2}e\sigma F-\Delta-B_\mu^2)s}
&=&e^{-i(10\lambda_{SO}^2+2\lambda_R^2)s}[\cos(\zeta s) +i\gamma^0
s_z\sin(\zeta s)]\{\cos (eBs)+\gamma^1\gamma^2\sin (eBs)+
\frac{1}{2}(1-\gamma^0s_z)[\cos(\xi s)-\cos(eBs)]\nonumber\\
&-&i\frac{12\lambda_{SO}\lambda_R \sin(\xi s)}{\xi}
(i\gamma^1s_y-\gamma^0\gamma^1s_x)
+\frac{1}{2}(1-\gamma^0s_z)\gamma^1\gamma^2[\frac{eB}{\xi}\sin(\xi
s)-\sin (eBs)]\},
\end{eqnarray}
\end{widetext}
where $\zeta=2\lambda_{SO}^2-6\lambda_R^2$,
$\xi=\sqrt{(24\lambda_{SO}\lambda_R)^2+(eB)^2}$. Substituting Eq.
(\ref{exp}) into Eq. (\ref{gxy}), applying Fourier transform in the
Matsubara representation and using the decomposition method over
Landau level poles\cite{chodos}\cite{gusy}, we can derive the
expression
\begin{equation}
\begin{array}{cll}
G(i\omega_m,\mathbf{k})=G_1(i\omega_m,\mathbf{k})
+G_2(i\omega_m,\mathbf{k})+G_3(i\omega_m,\mathbf{k})
\end{array}\label{eq35}
\end{equation}
after straightforward but somewhat complicated calculations. The
expressions of $G_i(i\omega_m,\mathbf{k})$ ($i$=1,2,3) are very
complicated and will be given in the appendix A. Whence, we can
further obtain the retarded and advanced Green's functions by the
analytic continuation
$G^{(R)}(\omega+i0,\mathbf{k})=G(i\omega_m\rightarrow\omega+i0,\mathbf{k})$
and
$G^{(A)}(\omega-i0,\mathbf{k})=G(i\omega_m\rightarrow\omega-i0,\mathbf{k})$.
When considering the influence of impurities, it is assumed that the
scattering rate $\Gamma$ on impurity is described phenomenologically
by a constant, and then the Green's functions acquire the form
\begin{widetext}
\begin{equation}
\begin{array}{cll}
G^{(R,A)}(\omega,\mathbf{k})&=&G_1^{(R,A)}(\omega\pm
i\Gamma,\mathbf{k}) +G_2^{(R,A)}(\omega\pm i\Gamma,\mathbf{k})
+G_3^{(R,A)}(\omega\pm i\Gamma,\mathbf{k}).
\end{array}\label{gra}
\end{equation}
\end{widetext}
In general, the scattering rate $\Gamma$, which is defined by
$\Gamma(\omega)=-Im\Sigma^R(\omega)$, is a frequency-dependent
quantity. It needs to be determined self-consistently from the
Schwinger-Dyson equations. The exact form of this equation
actually depends on the impurity scattering fashion, such as
short- or long-range scatterers. This kind of consideration have
been made for graphene in Ref. \cite{zheng2002prb}. But in this
paper, we mainly focus on the SOI effect on transport, and neglect
the exact form of interactions between impurities and electrons.

\section{Electronic conductivity}
The Kubo formula concerning the frequency-dependent electrical
conductivity as a linear response function to an external field
can be written as\cite{mahan.book}
\begin{equation}
\sigma_{ij}(\Omega)=\frac{\text{Im}\Pi_{ij}^R(\Omega+i0)}{\Omega},\label{sigij}
\end{equation}
where $i,j$ are the component indexes of coordinates, and
$\Pi_{ij}^R(\omega)$ is the retarded current-current correlation
function obtained by analytical continuation of the Matsubara
function
\begin{equation}
\Pi_{ij}(i\omega_n)=\frac{1}{V}\int_0^\beta d\tau
e^{i\omega_n\tau}\langle T_\tau J_i(\tau)J_j(0)\rangle,\\\
\omega_n=2\pi nT,\label{piij}
\end{equation}
where $V$ is the volume of the system, $\beta$ is the inverse
temperature, and $J_i(\tau)=\int d^2 r j_i(\tau,\mathbf{r})$ with
$j_i=-ev_F
\overline{\Psi}(\tau,\mathbf{r})\gamma^i\Psi(\tau,\mathbf{r})$.
Neglecting the impurity vertex corrections, the calculation of the
conductivity reduces to evaluation of the bubble diagram. Then Eq.
(\ref{sigij}) can be rewritten as
\begin{widetext}
\begin{eqnarray}
\sigma_{ij}(\Omega)&=&\frac{e^2v_F^2}{2\pi\Omega}
\text{Re}\int_{-\infty}^\infty d\omega \int\frac{d^2k}{(2\pi)^2}
tr\{[n_F(\omega)-n_F(\omega+\Omega)][\gamma^i
G^R(\omega+\Omega,\mathbf{k})
\gamma^j G^A(\omega,\mathbf{k})]\nonumber\\
&+&n_F(\omega+\Omega)\gamma^i G^A(\omega+\Omega,\mathbf{k})\gamma^j
G^A(\omega,\mathbf{k})-n_F(\omega)\gamma^i
G^R(\omega+\Omega,\mathbf{k})\gamma^jG^R(\omega,\mathbf{k})]\},
\end{eqnarray}
\end{widetext}
where $n_F(\omega)$ is the Fermi distribution function. Substituting
Eq. (\ref{gra}) into Eq. (14), we can obtain the longitudinal
conductivity
\begin{equation}
\begin{array}{cll}
\sigma_{xx}=\frac{2e^2v_F^2|eB|}{\pi^2}\text{Re}\int_{-\infty}^{\infty}
d\omega\frac{1}{4T\cosh^2\frac{\beta(\omega-\mu)}{2}} A_L(\omega)
\end{array}\label{sigxx}
\end{equation}
and the Hall conductivity
\begin{equation}
\begin{array}{cll}
\sigma_{xy}
=-\frac{2e^2v_F^2|eB|sgn(eB)}{\pi^2}\text{Im}\int_{-\infty}^{\infty}
d\omega\frac{1}{4T\cosh^2\frac{\beta(\omega-\mu)}{2}}A_H(\omega),
\end{array}\label{sigxy}
\end{equation}
where all the quantities on the right-hand side are calculated in
the Appendix B. Eqs. (\ref{sigxx}) and (\ref{sigxy}) establish the
fundamental basis for investigating the SOI effect on the quantum
transport properties of the Dirac fermions in graphene.

In the limit of zero field, the Hall conductivity becomes zero.
While for the longitudinal conductivity, using the asymptotic
expansions
\begin{equation}
\psi(z)=\ln
z-\frac{1}{2z}-\frac{1}{12z^2}+\frac{1}{120z^4}+O(\frac{1}{z^5}),
\end{equation}
we arrive at
\begin{widetext}
\begin{eqnarray}
\begin{array}{cll}
A_L(\omega)&=&[\frac{(\omega^2-\Gamma^2)}{6\lambda_R^2}
-\frac{6\lambda_R^2(\omega^2+\Gamma^2)}{(2\Gamma\omega)^2+(6\lambda_R^2)^2}
]\ln\frac{(-4\lambda_R^2-\omega^2+
\Gamma^2)^2+(2\Gamma\omega)^2}{(8\lambda_R^2-\omega^2+
\Gamma^2)^2+(2\Gamma\omega)^2}
+\frac{\Gamma\omega(\omega^2+\Gamma^2)}{(\Gamma\omega)^2+(3\lambda_R^2)^2}
[\arctan\frac{\omega+2\sqrt{2}\lambda_{R}}{\Gamma}+\arctan\frac{\omega-2\sqrt{2}\lambda_{R}}{\Gamma}
\\
&+&\arctan\frac{\omega+2\lambda_{R}}{\Gamma}\arctan\frac{\omega-2\lambda_{R}}{\Gamma}]-\frac{2\Gamma\omega}{3\lambda_R^2}
[\arctan\frac{\omega+2\sqrt{2}\lambda_{R}}{\Gamma}+\arctan\frac{\omega-2\sqrt{2}\lambda_{R}}{\Gamma}
-\arctan\frac{\omega+2\lambda_{R}}{\Gamma}-\arctan\frac{\omega-2\lambda_{R}}{\Gamma}]
\end{array}
\end{eqnarray}
\end{widetext}
for $\lambda_{SO}=0$. When $T\rightarrow0$, $|\mu|>>\lambda_R$,
$\Gamma$, the longitudinal conductivity can be further expressed as
\begin{equation}
\sigma_{xx} =\frac{4e^2}{\pi}
\frac{|\mu|/\Gamma}{1+(3\lambda_R^2/\Gamma\mu)^2}.\label{sigxxlm}
\end{equation}

\section{Results and Discussion}
To investigate numerically the behavior of electrical conductivity,
we need to restore the whole model parameters in Eqs. (\ref{sigxx})
and (\ref{sigxy}). Thus, one should carry out the replacements:
\begin{figure}[ht]
\centering{
\includegraphics[width=8cm]{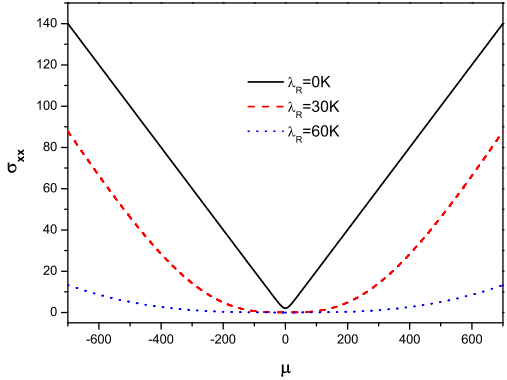}}
\caption{The longitudinal conductivity $\sigma_{xx}$ measured in
$2e^2/h$ units as a function of the chemical potential $\mu$ for the
different values of $\lambda_{R}$. We take $B = 0 \mbox{T}$, $T =
3\mbox{K}$, $\Gamma = 5 \mbox{K}$, and
$\lambda_{SO}=0.001\mbox{K}$.} \label{fig:3-1}
\end{figure}
$T\rightarrow k_BT, eB\rightarrow \hbar eBv_F^2$. In the following,
we mainly discuss the Rashba SOI effect on the transport properties
since the intrinsic SOI is very small, while the Rashba SOI can be
tunable by a perpendicular electric field. In Figs. 1 and 2, we show
the chemical potential $\mu$ dependence of the longitudinal
conductivity for the different Rashba spin orbit parameter
$\lambda_R$ at zero or nonzero field. For zero field (see Fig. 1),
one can see that when $\lambda_R=0$, the conductivity is
\begin{figure}[ht]
\centering{
\includegraphics[width=8cm]{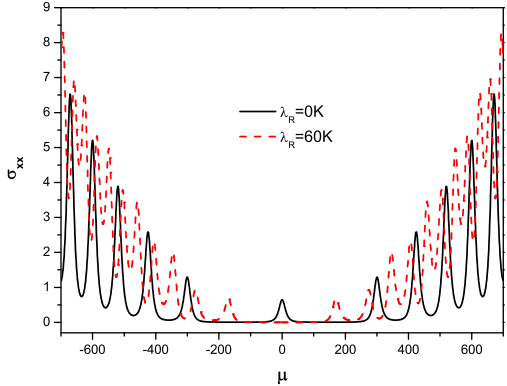}}
\caption{The longitudinal conductivity measured in $2e^2/h$ units as
a function of the chemical potential $\mu$ for the different values
of $\lambda_{R}$ at $B=1\mbox{T}$. The other parameters are taken
the same as Fig. \ref{fig:3-1}}\label{fig:3-2}
\end{figure}
proportional to $|\mu|$ and tends to the known quantum-limited
minimal value $4e^2/h$ at zero chemical potential\cite{novoselov2}.
For $\lambda_R\neq 0$, there exists a threshold chemical potential
$\mu_c$ which increases with increasing $\lambda_R$. When the
chemical potential is smaller than $\mu_c$, the longitudinal
conductivity becomes almost independent of $\mu$; while for
$\mu>\mu_c$, the $\sigma_{xx}$-$\mu$ curves recover the linear
relation. This tendency agrees with Eq. (\ref{sigxxlm}). For nonzero
field case in Fig. 2, we observe SdH oscillations of the
conductivity due to the Landau-level crossing of the Fermi
level\cite{sharapov2003prb}. From Fig. 2, It is clearly seen that
when $\lambda_R\neq 0$, each oscillation peak is split into two
implicit peaks, and the splitting peaks shift by $\lambda_R$. This
is due to the spin-orbit splitting of the Landau levels.
\begin{figure}[ht]
\centering{
\includegraphics[width=8cm]{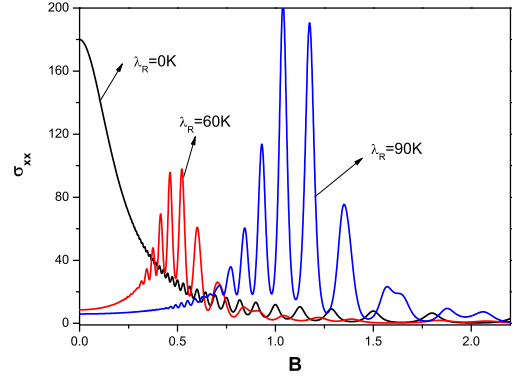}}
\caption{The magnetic field dependence of the longitudinal
conductivity measured in $2e^2/h$ units for the different values of
$\lambda_{R}$ at $\mu=-600\mbox{K}$. The other parameters are taken
as Fig. \ref{fig:3-1}.}\label{fig:3-3}
\end{figure}

The longitudinal conductivity as a function of the magnetic field B
for the different $\lambda_R$ is shown in Fig. 3. For $\lambda_R=0$,
the longitudinal conductivity decreases and intervals between the
neighboring SdH oscillation peaks become large with increasing B,
which reflects the fact that in the presence of the magnetic field
only the transitions between neighboring Landau levels contribute to
electrical conductivity, while a further increase of the magnetic
field leads to increasing of the distance between neighboring Landau
level, thus suppresses the transitions between them. When $B$ is
large enough, the conductivity becomes independent of $B$ since the
lowest Landau level is filled which is always below the Fermi level.
These observations are quite consistent with the previous
studies\cite{sharapov2003prb}\cite{gusynin2}. In particular, it is
interesting to note that when the Rashba SOI presents, the
longitudinal conductivity exhibits the characteristic feature that
the SdH oscillations are enhanced largely at certain positions,
however damped at other positions. Such SdH as a beating pattern
have been observed in two dimensional electron gas\cite{luoprb1990}.
From Fig. 3, one can find that the enhanced positions and amplitudes
of the SdH oscillations can be tuned by the Rashba SOI due to shift
of one set of Landau level by $\lambda_R$.
\begin{figure}[ht]
\centering{
\includegraphics[width=8cm]{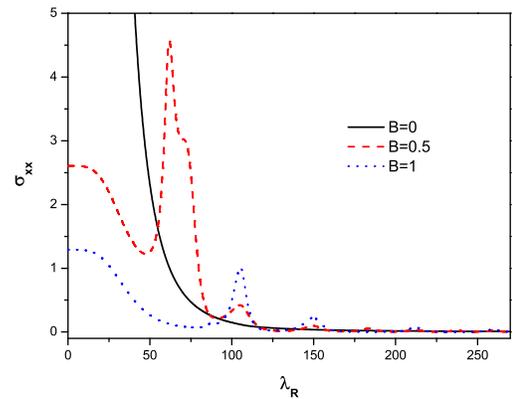}}
\caption{The longitudinal conductivity measured in $2e^2/h$ units as
a function of $\lambda_R$ for the different magnetic field $B$ at
$\mu=-300\mbox{K}$. The other parameters are taken the same as Fig.
\ref{fig:3-1}. }\label{fig:3-4}
\end{figure}

Figure 4 shows the longitudinal conductivity versus the Rashba SOI
parameter $\lambda_R$ for the different magnetic field $B$. It is
found that when the magnetic field is applied, the longitudinal
conductivity as a function of $\lambda_R$ behaves as the
oscillation. It is because the Rashba SOI leads to the shift of
landau level, the longitudinal conductivity shows a maximum each
time a Landau level passes through the Fermi level of system, and a
minimum when the Fermi level is situated between two Landau levels.
\begin{figure}[ht]
\centering{
\includegraphics[width=8cm]{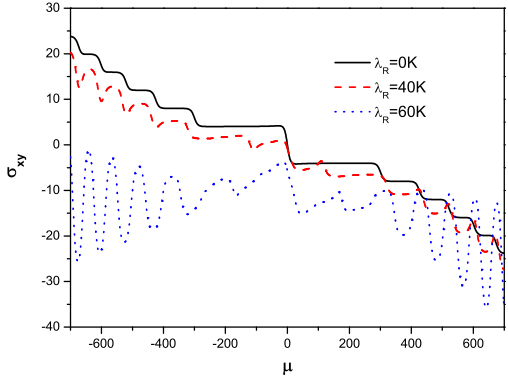}}
\caption{The Hall conductivity measured in $2e^2/h$ units as a
function of the chemical potential $\mu$ for the different values of
$\lambda_{R}$ at $B=1\mbox{T}$. The other parameters are taken the
same as Fig. \ref{fig:3-1} }\label{fig:3-5}
\end{figure}

Figure 5 shows Hall conductivity as a function of the chemical
potential $\mu$ for the different $\lambda_R$.  When $\lambda_R=0$,
the Hall conductivity has a steplike structure as a function of
$\mu$, which reflect the quantum Hall effect. While the Rashba SOI
opens, the Hall steps become narrow and the step near $\mu=0$ is
split into two steps. It is observed that the Hall conductivity
displays peaks instead of a plateau at larger $\lambda_R$. There is
no Hall plateau in the cases of sufficiently strong Rashba SOI. This
result suggests that the Rashba SOI tends to suppress the quantum
Hall effect in graphene.

\section{SUMMARY}
We have investigated the effect of the SOIs on transport of Dirac
fermions in graphene on the basis of amplitude estimation of Yao
\emph{et al}. and Min \emph{et al}. for the SOIs. Using the
Schwinger proper-time method, decomposition over Landau level poles
and Kubo formula, we obtain the analytical expressions for both
longitudinal and Hall conductivities. It has been found that when
the Rashba SOI is applied, the longitudinal conductivity versus the
chemical potential deviates the linear relation at zero magnetic
field. For nonzero magnetic field, the SdH oscillation in the
longitudinal conductivity is observed, and each SdH oscillation peak
is split into two peaks as the Rashba SOI is applied. While the
oscillation in the longitudinal conductivity as a function of the
magnetic field exhibits a beating pattern with the Rashba SOI turned
on. It is also shown that the Rashba SOI tends to suppress the
quantum Hall effect in graphene.

\begin{acknowledgments}
Ding gratefully acknowledge financial support from Changsha
University of Science and Technology, and Education Department of
Hunan Province, China. The work of Zhou was supported by National
Natural Science Foundation of China (Grant No. 10574042) and
Specialized Research Fund for the Doctoral Program of Higher
Education of China (Grant No. 20060542002).
\end{acknowledgments}

\appendix
\section{}
\label{sec:A} The Green functions $G_i^{(R,A)}$ in Eq. (11) are
given by
\begin{widetext}
\begin{equation}
\begin{array}{cll}
G_1^{(R,A)}(\omega)&=&(H_{s}+\gamma^0(\omega\pm
i\Gamma))[A_1^{(R,A)}(\omega)-i\gamma^1\gamma^2
sgn(eB)A_2^{(R,A)}(\omega)]
-i(\gamma^1k_2-\gamma^2k_1)sgn(eB)[B_1^{(R,A)}(\omega)
\\
&-&A_2^{(R,A)}(\omega)]+(\gamma^1
k_1+\gamma^2k_2)[B_2^{(R,A)}(\omega)-A_1^{(R,A)}(\omega)],
\end{array}
\end{equation}
\begin{equation}
\begin{array}{cll}
G_2^{(R,A)}(\omega)&=&(H_{s}+\gamma^0(\omega\pm
i\Gamma))\gamma^0s_z[A_3^{(R,A)}(\omega)-i\gamma^1\gamma^2
sgn(eB)A_4^{(R,A)}(\omega)]
-i(\gamma^1k_2-\gamma^2k_1)\gamma^0s_zsgn(eB)[B_3^{(R,A)}(\omega)
\\
&-&A_4^{(R,A)}(\omega)]+(\gamma^1 k_1+\gamma^2
k_2)\gamma^0s_z[B_4^{(R,A)}(\omega)-A_3^{(R,A)}(\omega)],
\end{array}
\end{equation}
\begin{eqnarray}
G_3^{(R,A)}(\omega)&=&-\frac{12\lambda_{SO}\lambda_R
}{\xi}(H_{s}+\gamma^0(\omega\pm
i\Gamma))(i\gamma^1s_y-\gamma^0\gamma^1s_x)(I_4^{(R,A)}(\omega)+\gamma^0s_z
I_8^{(R,A)}(\omega))+\frac{12\lambda_{SO}\lambda_R
}{\xi}(\gamma^1k_1\nonumber\\
&+&\gamma^2k^2)(I_4^{(R,A)}(\omega)+i\gamma^1\gamma^2sgn(eB)J_4^{(R,A)}(\omega))
(i\gamma^1s_y-\gamma^0\gamma^1s_x)+\frac{12\lambda_{SO}\lambda_R}
{\xi}(\gamma^1k_1+\gamma^2k^2)(I_8^{(R,A)}(\omega)\nonumber\\
&+&i\gamma^1\gamma^2sgn(eB)J_8^{(R,A)}(\omega))
(i\gamma^1s_y-\gamma^0\gamma^1s_x)\gamma^0s_z,
\end{eqnarray}
where $A_1=I_1+I_3+I_5-I_7$, $A_2=I_2+I_4/\chi+I_6-I_8/\chi$,
$A_3=I_1-I_3+I_5+I_7$, $A_4=I_2-I_4/\chi+I_6+I_8/\chi$,
$B_1=I_2+J_2+I_6-J_5$, $B_2=J_1+J_3/\chi+J_4-J_6/\chi$,
$B_3=I_2-J_2+I_6+J_5$, $B_4=J_1-J_3/\chi+J_4+J_6/\chi$, in which
\begin{equation}
\begin{array}{cll}
I_1^{(R,A)}(\omega)
&=&\frac{1}{2}e^{-\alpha}\sum\limits_{n=0}^\infty
(-1)^n[\frac{L_n(2\alpha)-L_{n-1}(2\alpha)}{(\omega\pm
i\Gamma)^2+10\lambda_{SO}^2+2\lambda_R^2-\zeta+2n|eB|}
+\frac{L_n(2\alpha)-L_{n-1}(2\alpha)}{(\omega\pm
i\Gamma)^2+10\lambda_{SO}^2+2\lambda_R^2+\zeta+2n|eB|}],
\end{array}
\end{equation}
\begin{equation}
\begin{array}{cll}
I_2^{(R,A)}(\omega)
&=&\frac{1}{2}e^{-\alpha}\sum\limits_{n=0}^\infty
(-1)^n[\frac{L_n(2\alpha)+L_{n-1}(2\alpha)}{(\omega\pm
i\Gamma)^2+10\lambda_{SO}^2+2\lambda_R^2-\zeta+2n|eB|}
+\frac{L_n(2\alpha)+L_{n-1}(2\alpha)}{(\omega\pm
i\Gamma)^2+10\lambda_{SO}^2+2\lambda_R^2+\zeta+2n|eB|}],
\end{array}
\end{equation}
\begin{equation}
\begin{array}{cll}
I_3^{(R,A)}(\omega)
&=&\frac{1}{2}e^{-\alpha}\sum\limits_{n=0}^\infty (-1)^n
[\frac{L_n(2\alpha)}{(\omega\pm
i\Gamma)^2+10\lambda_{SO}^2+2\lambda_R^2 -\zeta-\xi+(2n+1)|eB|}
+\frac{L_n(2\alpha)}{(\omega\pm
i\Gamma)^2+10\lambda_{SO}^2+2\lambda_R^2+\zeta+\xi+(2n+1)|eB|}\\
&&+\frac{L_n(2\alpha)}{(\omega\pm
i\Gamma)^2+10\lambda_{SO}^2+2\lambda_R^2-\zeta+\xi+(2n+1)|eB|}
+\frac{L_n(2\alpha)}{(\omega\pm
i\Gamma)^2+10\lambda_{SO}^2+2\lambda_R^2+\zeta-\xi+(2n+1)|eB|}],
\end{array}
\end{equation}
\begin{equation}
\begin{array}{cll}
I_4^{(R,A)}(\omega)
&=&\frac{1}{2}e^{-\alpha}\sum\limits_{n=0}^\infty (-1)^n
[\frac{L_n(2\alpha)}{(\omega\pm
i\Gamma)^2+10\lambda_{SO}^2+2\lambda_R^2 -\zeta-\xi+(2n+1)|eB|}
-\frac{L_n(2\alpha)}{(\omega\pm
i\Gamma)^2+10\lambda_{SO}^2+2\lambda_R^2+\zeta
+\xi+(2n+1)|eB|}\\
&& -\frac{L_n(2\alpha)}{(\omega\pm
i\Gamma)^2+10\lambda_{SO}^2+2\lambda_R^2-\zeta+\xi+(2n+1)|eB|}
+\frac{L_n(2\alpha)}{(\omega\pm
i\Gamma)^2+10\lambda_{SO}^2+2\lambda_R^2+\zeta-\xi+(2n+1)|eB|}],
\end{array}
\end{equation}
\begin{equation}
\begin{array}{cll}
I_5^{(R,A)}(\omega)
&=&\frac{1}{2}e^{-\alpha}\sum\limits_{n=0}^\infty (-1)^n
[\frac{L_n(2\alpha)-L_{n-1}(2\alpha)}{(\omega\pm
i\Gamma)^2+10\lambda_{SO}^2+2\lambda_R^2-\zeta+2n|eB|}
-\frac{L_n(2\alpha)-L_{n-1}(2\alpha)}{(\omega\pm
i\Gamma)^2+10\lambda_{SO}^2+2\lambda_R^2+\zeta+2n|eB|}],
\end{array}
\end{equation}
\begin{equation}
\begin{array}{cll}
I_6^{(R,A)}(\omega)
&=&\frac{1}{2}e^{-\alpha}\sum\limits_{n=0}^\infty
(-1)^n[\frac{L_n(2\alpha)+L_{n-1}(2\alpha)}{(\omega\pm
i\Gamma)^2+10\lambda_{SO}^2+2\lambda_R^2-\zeta+2n|eB|}
-\frac{L_n(2\alpha)+L_{n-1}(2\alpha)}{(\omega\pm
i\Gamma)^2+10\lambda_{SO}^2+2\lambda_R^2+\zeta+2n|eB|}],
\end{array}
\end{equation}
\begin{equation}
\begin{array}{cll}
I_7^{(R,A)}(\omega)
&=&\frac{1}{2}e^{-\alpha}\sum\limits_{n=0}^\infty (-1)^n
[\frac{L_n(2\alpha)}{(\omega\pm i\Gamma)^2+10\lambda_{SO}^2
+2\lambda_R^2-\zeta-\xi+(2n+1)|eB|} -\frac{L_n(2\alpha)}{(\omega\pm
i\Gamma)^2+10\lambda_{SO}^2+2\lambda_R^2+\zeta+\xi+(2n+1)|eB|}\\
&&+\frac{L_n(2\alpha)}{(\omega\pm
i\Gamma)^2+10\lambda_{SO}^2+2\lambda_R^2-\zeta+\xi+(2n+1)|eB|}
-\frac{L_n(2\alpha)}{(\omega\pm
i\Gamma)^2+10\lambda_{SO}^2+2\lambda_R^2+\zeta-\xi+(2n+1)|eB|}],
\end{array}
\end{equation}
\begin{equation}
\begin{array}{cll}
I_8^{(R,A)}(\omega)
&=&\frac{1}{2}e^{-\alpha}\sum\limits_{n=0}^\infty (-1)^n
[\frac{L_n(2\alpha)}{(\omega\pm i\Gamma)^2
+10\lambda_{SO}^2+2\lambda_R^2-\zeta-\xi+(2n+1)|eB|}
+\frac{L_n(2\alpha)}{(\omega\pm
i\Gamma)^2+10\lambda_{SO}^2+2\lambda_R^2+\zeta+\xi+(2n+1)|eB|}\\
&& -\frac{L_n(2\alpha)}{(\omega\pm
i\Gamma)^2+10\lambda_{SO}^2+2\lambda_R^2-\zeta+\xi+(2n+1)|eB|}
-\frac{L_n(2\alpha)}{(\omega\pm
i\Gamma)^2+10\lambda_{SO}^2+2\lambda_R^2+\zeta-\xi+(2n+1)|eB|}],
\end{array}
\end{equation}
\begin{equation}
\begin{array}{cll}
J_1^{(R,A)}(\omega)
&=&\frac{1}{2}e^{-\alpha}\sum\limits_{n=0}^\infty
(-1)^n[\frac{4L_{n-1}^1(2\alpha)+L_n(2\alpha)-L_{n-1}(2\alpha)}{(\omega\pm
i\Gamma)^2+10\lambda_{SO}^2+2\lambda_R^2-\zeta+2n|eB|}
+\frac{4L_{n-1}^1(2\alpha)+L_n(2\alpha)-L_{n-1}(2\alpha)}{(\omega\pm
i\Gamma)^2+10\lambda_{SO}^2+2\lambda_R^2+\zeta+2n|eB|}],
\end{array}
\end{equation}
\begin{equation}
\begin{array}{cll}
J_2^{(R,A)}(\omega)
&=&\frac{1}{2}e^{-\alpha}\sum\limits_{n=0}^\infty (-1)^n
[\frac{L_n^1(2\alpha)+L_{n-1}^1(2\alpha)}{(\omega\pm
i\Gamma)^2+10\lambda_{SO}^2 +2\lambda_R^2-\zeta-\xi+(2n+1)|eB|}
+\frac{L_n^1(2\alpha)+L_{n-1}^1(2\alpha)}{(\omega\pm
i\Gamma)^2+10\lambda_{SO}^2+2\lambda_R^2+\zeta+\xi+(2n+1)|eB|}\\
&&+\frac{L_n^1(2\alpha)+L_{n-1}^1(2\alpha)}{(\omega\pm
i\Gamma)^2+10\lambda_{SO}^2+2\lambda_R^2-\zeta+\xi+(2n+1)|eB|}
+\frac{L_n^1(2\alpha)+L_{n-1}^1(2\alpha)} {(\omega\pm
i\Gamma)^2+10\lambda_{SO}^2+2\lambda_R^2+\zeta-\xi+(2n+1)|eB|}],
\end{array}
\end{equation}
\begin{equation}
\begin{array}{cll}
J_3^{(R,A)}(\omega)
&=&\frac{1}{2}e^{-\alpha}\sum\limits_{n=0}^\infty (-1)^n
[\frac{L_n^1(2\alpha)+L_{n-1}^1(2\alpha)}{(\omega\pm i\Gamma)^2
+10\lambda_{SO}^2+2\lambda_R^2-\zeta-\xi+(2n+1)|eB|}
-\frac{L_n^1(2\alpha)+L_{n-1}^1(2\alpha)}{(\omega\pm
i\Gamma)^2+10\lambda_{SO}^2+2\lambda_R^2+\zeta+\xi+(2n+1)|eB|}\\
&&-\frac{L_n^1(2\alpha)+L_{n-1}^1(2\alpha)}{(\omega\pm
i\Gamma)^2+10\lambda_{SO}^2+2\lambda_R^2-\zeta+\xi+(2n+1)|eB|}
+\frac{L_n^1(2\alpha)+L_{n-1}^1(2\alpha)}{(\omega\pm i\Gamma)^2
+10\lambda_{SO}^2+2\lambda_R^2+\zeta-\xi+(2n+1)|eB|}],
\end{array}
\end{equation}
\begin{equation}
\begin{array}{cll}
J_4^{(R,A)}(\omega)
&=&\frac{1}{2}e^{-\alpha}\sum\limits_{n=0}^\infty
(-1)^n[\frac{4L_{n-1}^1(2\alpha)+L_n(2\alpha)-L_{n-1}(2\alpha)}{(\omega\pm
i\Gamma)^2+10\lambda_{SO}^2+2\lambda_R^2-\zeta+2n|eB|}
-\frac{4L_{n-1}^1(2\alpha)+L_n(2\alpha)-L_{n-1}(2\alpha)}{(\omega\pm
i\Gamma)^2+10\lambda_{SO}^2+2\lambda_R^2+\zeta+2n|eB|}],
\end{array}
\end{equation}
\begin{equation}
\begin{array}{cll}
J_5^{(R,A)}(\omega)
&=&\frac{1}{2}e^{-\alpha}\sum\limits_{n=0}^\infty (-1)^n
[\frac{L_n^1(2\alpha)+L_{n-1}^1(2\alpha)}{(\omega\pm i\Gamma)^2
+10\lambda_{SO}^2+2\lambda_R^2-\zeta-\xi+(2n+1)|eB|}
-\frac{L_n^1(2\alpha)+L_{n-1}^1(2\alpha)}{(\omega\pm
i\Gamma)^2+10\lambda_{SO}^2+2\lambda_R^2+\zeta+\xi+(2n+1)|eB|}\\
&&+\frac{L_n^1(2\alpha)+L_{n-1}^1(2\alpha)}{(\omega\pm
i\Gamma)^2+10\lambda_{SO}^2+2\lambda_R^2-\zeta+\xi+(2n+1)|eB|}
-\frac{L_n^1(2\alpha)+L_{n-1}^1(2\alpha)}{(\omega\pm i\Gamma)^2
+10\lambda_{SO}^2+2\lambda_R^2+\zeta-\xi+(2n+1)|eB|}],
\end{array}
\end{equation}
\begin{equation}
\begin{array}{cll}
J_6^{(R,A)}(\omega)
&=&\frac{1}{2}e^{-\alpha}\sum\limits_{n=0}^\infty (-1)^n
[\frac{L_n^1(2\alpha)+L_{n-1}^1(2\alpha)}{(\omega\pm i\Gamma)^2
+10\lambda_{SO}^2+2\lambda_R^2-\zeta-\xi+(2n+1)|eB|}
-\frac{L_n^1(2\alpha)+L_{n-1}^1(2\alpha)}{(\omega\pm
i\Gamma)^2+10\lambda_{SO}^2+2\lambda_R^2+\zeta+\xi+(2n+1)|eB|}\\
&&+\frac{L_n^1(2\alpha)+L_{n-1}^1(2\alpha)}{(\omega\pm
i\Gamma)^2+10\lambda_{SO}^2+2\lambda_R^2-\zeta+\xi+(2n+1)|eB|}
-\frac{L_n^1(2\alpha)+L_{n-1}^1(2\alpha)}{(\omega\pm i\Gamma)^2
+10\lambda_{SO}^2+2\lambda_R^2+\zeta-\xi+(2n+1)|eB|}],
\end{array}
\end{equation}
with $L_n^i(z)$ being the generalized Laguerre polynomials,
$\alpha=-\mathbf{k}^2/|eB|,$ and $\chi=\xi/|eB|$.

\section{}
\label{sec:B} Substituting Eq. (\ref{gra}) into the trace Eq. (14),
then it is evaluated after a somewhat tedious calculation
\begin{eqnarray}
tr\{\gamma^i S^{(R,A)}(\omega',\mathbf{k})\gamma^j
S^{(R,A)}(\omega,\mathbf{k})\}&=&\delta_{ij}(\omega'\pm
i\Gamma)(\omega\pm
i\Gamma)[A_1^{(R,A)}(\omega')A_1^{(R,A)}(\omega)-A_2^R(\omega')A_2^A(\omega)
+A_4^{(R,A)}(\omega')A_4^{(R,A)}(\omega)\nonumber\\
&-&A_3^{(R,A)}(\omega')A_3^{(R,A)}(\omega)]
+i\epsilon_{ij}sgn(eB)(\omega'\pm i\Gamma)(\omega\pm
i\Gamma)[A_2^{(R,A)}(\omega')A_1^{(R,A)}(\omega)\nonumber\\
&-&A_1^{(R,A)}(\omega')A_2^{(R,A)}(\omega)
+A_3^{(R,A)}(\omega')A_4^{(R,A)}(\omega)
-A_4^{(R,A)}(\omega')A_3^{(R,A)}(\omega)]\nonumber\\
&+&(2k_ik_j-\delta_{ij}\mathbf{k}^2)[(B_1^{(R,A)}(\omega')
-A_2^{(R,A)}(\omega'))(B_1^{(R,A)}(\omega)
-A_2^{(R,A)}(\omega))\nonumber\\
&+&(B_2^{(R,A)}(\omega')
-A_1^{(R,A)}(\omega'))(B_2^{(R,A)}(\omega)-A_1^{(R,A)}(\omega))
+(B_3^{(R,A)}(\omega') -A_4^{(R,A)}(\omega'))\nonumber\\
&\times&(B_3^{(R,A)}(\omega) -A_4^{(R,A)}(\omega))
+(B_4^{(R,A)}(\omega')
-A_3^{(R,A)}(\omega'))(B_4^{(R,A)}(\omega)-A_3^{(R,A)}(\omega))]\nonumber\\
&-&isgn(eB)[\delta_{ij}(-1)^{i-1}2k_1k_2+\epsilon_{ij}(k_j^2-k_i^2)][(B_1^{(R,A)}(\omega')
-A_2^{(R,A)}(\omega'))
(B_2^{(R,A)}(\omega)-A_1^{(R,A)}(\omega))\nonumber\\
&+&(B_3^{(R,A)}(\omega')
-A_4^{(R,A)}(\omega'))(B_4^{(R,A)}(\omega)-A_3^{(R,A)}(\omega))
+(B_2^{(R,A)}(\omega')-A_1^{(R,A)}(\omega'))\nonumber\\
&\times&(B_1^{(R,A)}(\omega) -A_2^{(R,A)}(\omega))
+(B_4^{(R,A)}(\omega')-A_3^{(R,A)}(\omega'))(B_3^{(R,A)}(\omega) -A_4^{(R,A)}(\omega))]\nonumber\\
&+&\delta_{ij}2(\frac{12\lambda_{SO}\lambda_R
}{\xi})^2(I_4^{(R,A)}(\omega')I_4^{(R,A)}(\omega)-J_4^{(R,A)}(\omega')J_4
-I_8^{(R,A)}(\omega')I_8^{(R,A)}(\omega)\nonumber\\
&+&J_8^{(R,A)}(\omega')J_8^{(R,A)}(\omega))\mathbf{k}^2
-i\epsilon_{ij}sgn(eB)2(\frac{12\lambda_{SO}\lambda_R
}{\xi})^2[I_4^{(R,A)}(\omega')J_4^{(R,A)}(\omega)\nonumber\\
&-&J_4^{(R,A)}(\omega')I_4^{(R,A)}(\omega)
-I_8^{(R,A)}(\omega')J_8^{(R,A)}(\omega)+J_8^{(R,A)}(\omega')
I_8^{(R,A)}(\omega)]\mathbf{k}^2,
\end{eqnarray}
where $\epsilon_{ij}$ is antisymmetric tensor ($\epsilon_{12}=1$).
Integrating over momenta in Eq. (14), we obtain the longitudinal
conductivity
\begin{equation}
\begin{array}{cll}
\sigma_{xx}
=\sigma_{xx}(\Omega\rightarrow0)=\frac{2e^2v_F^2|eB|}{\pi^2}Re\int_{-\infty}^{\infty}
d\omega\frac{1}{4T\cosh^2\frac{\beta(\omega-\mu)}{2}}\{X_L
+\frac{1}{2|eB|}Y_L-(\frac{6\lambda_{SO}\lambda_R}{\xi})^2Z_L\},
\end{array}\label{sigxxa1}
\end{equation}
where
\begin{eqnarray}
\begin{array}{cll}
X_L&=&\frac{1}{(\omega+i\Gamma)^2-(10\lambda_{SO}^2+2\lambda_R^2)
-\zeta-\xi+|eB|}[\frac{2(\omega^2+\Gamma^2)}{i4\Gamma\omega-2\zeta-\xi+|eB|-2|eB|}
-\frac{2(\omega+i\Gamma)^2}{-2\zeta-\xi+|eB|-2|eB|}]
+\frac{(1-\frac{1}{\chi})}
{(\omega+i\Gamma)^2-(10\lambda_{SO}^2+2\lambda_R^2)
-\zeta+\xi+|eB|}\\
&\times&[\frac{(\omega^2+\Gamma^2)}{i4\Gamma\omega-2\zeta+\xi+|eB|-2|eB|}
-\frac{(\omega+i\Gamma)^2}{-2\zeta+\xi+|eB|-2|eB|}]-\frac{1}
{(\omega+i\Gamma)^2-(10\lambda_{SO}^2+2\lambda_R^2)+\zeta}
[\frac{(1-\frac{1}{\chi})(\omega^2+\Gamma^2)}{-i4\Gamma\omega-2\zeta-\xi-|eB|+2|eB|}
\\
&-&\frac{(1-\frac{1}{\chi})(\omega+i\Gamma)^2}{-2\zeta-\xi-|eB|+2|eB|}
+\frac{2(\omega^2+\Gamma^2)}{-i4\Gamma\omega-2\zeta+\xi-|eB|+2|eB|}
-\frac{2(\omega+i\Gamma)^2}{-2\zeta+\xi-|eB|+2|eB|}],
\end{array}
\end{eqnarray}
\begin{equation}
\begin{array}{cll}
Y_L&=&w_1\psi(-\frac{(\omega+i\Gamma)^2-(10\lambda_{SO}^2+2\lambda_R^2)
+\zeta}{2|eB|})+w_2\psi(-\frac{(\omega+i\Gamma)^2-(10\lambda_{SO}^2+2\lambda_R^2)
-\zeta-\xi+|eB|}{2|eB|})
+w_3\psi(-\frac{(\omega+i\Gamma)^2-(10\lambda_{SO}^2+2\lambda_R^2)
-\zeta+\xi+|eB|}{2|eB|})\\
&+&w_4\psi(-\frac{(\omega+i\Gamma)^2-(10\lambda_{SO}^2+2\lambda_R^2)
-\zeta+\xi-|eB|}{2|eB|})
+w_5\psi(-\frac{(\omega+i\Gamma)^2-(10\lambda_{SO}^2+2\lambda_R^2)
-\zeta-\xi-|eB|}{2|eB|}),
\end{array}
\end{equation}
\begin{equation}
\begin{array}{cll}
Z_L &=&z_1\psi(-\frac{(\omega+i\Gamma)^2-(10\lambda_{SO}^2
+2\lambda_R^2)+\zeta+\xi-|eB|}{2|eB|}-z_2
\psi(-\frac{(\omega+i\Gamma)^2-(10\lambda_{SO}^2+2\lambda_R^2)-\zeta-\xi+|eB|}{2|eB|})
+z_3\psi(-\frac{(\omega+i\Gamma)^2-(10\lambda_{SO}^2+2\lambda_R^2)-\zeta+\xi+|eB|}{2|eB|})\\
&-&z_4\psi(-\frac{(\omega+i\Gamma)^2-(10\lambda_{SO}^2+2\lambda_R^2)+\zeta-\xi-|eB|}{2|eB|})
+z_5\psi(-\frac{(\omega+i\Gamma)^2-(10\lambda_{SO}^2+2\lambda_R^2)-\zeta-\xi-|eB|}{2|eB|})
-z_6\psi(-\frac{(\omega+i\Gamma)^2-(10\lambda_{SO}^2+2\lambda_R^2)+\zeta+\xi+|eB|}{2|eB|})\\
&+&z_7\psi(-\frac{(\omega+i\Gamma)^2-(10\lambda_{SO}^2+2\lambda_R^2)+\zeta-\xi+|eB|}{2|eB|})
-z_8\psi(-\frac{(\omega+i\Gamma)^2-(10\lambda_{SO}^2+2\lambda_R^2)-\zeta+\xi-|eB|}{2|eB|}),
\end{array}
\end{equation}
with $\psi(z)$ being the digamma function and
$$
\begin{array}{cll}
w_1
&=&\frac{2(\omega^2+\Gamma^2)}{-i4\Gamma\omega-2\zeta-\xi+|eB|-2|eB|}
-\frac{2(\omega+i\Gamma)^2}{-2\zeta-\xi+|eB|-2|eB|}
+\frac{(1-\frac{1}{\chi})(\omega^2+\Gamma^2)}{-i4\Gamma\omega-2\zeta+\xi+|eB|-2|eB|}
-\frac{(1-\frac{1}{\chi})(\omega+i\Gamma)^2}{-2\zeta+\xi+|eB|-2|eB|}\\
&&+\frac{(1-\frac{1}{\chi})(\omega^2+\Gamma^2)}{-i4\Gamma\omega-2\zeta-\xi-|eB|+2|eB|}
-\frac{(1-\frac{1}{\chi})(\omega+i\Gamma)^2}{-2\zeta-\xi-|eB|+2|eB|}
+\frac{2(\omega^2+\Gamma^2)}{-i4\Gamma\omega-2\zeta+\xi-|eB|+2|eB|}
-\frac{2(\omega+i\Gamma)^2}{-2\zeta+\xi-|eB|+2|eB|},
\end{array}
$$
$$
\begin{array}{cll}
w_2&=&\frac{2(\omega^2+\Gamma^2)}{-i4\Gamma\omega+2\zeta+\xi-|eB|+2|eB|}
-\frac{2(\omega+i\Gamma)^2}{2\zeta+\xi-|eB|+2|eB|},\ \ w_3
=\frac{(1-\frac{1}{\chi})(\omega^2+\Gamma^2)}{-i4\Gamma\omega+2\zeta-\xi-|eB|+2|eB|}
-\frac{(1-\frac{1}{\chi})(\omega+i\Gamma)^2}{2\zeta-\xi-|eB|+2|eB|},
\end{array}
$$
$$
\begin{array}{cll}
w_4
=\frac{2(\omega^2+\Gamma^2)}{-i4\Gamma\omega+2\zeta-\xi+|eB|-2|eB|}
-\frac{2(\omega+i\Gamma)^2}{2\zeta-\xi+|eB|-2|eB|},\ \ w_5=
\frac{(1-\frac{1}{\chi})(\omega^2+\Gamma^2)}{-i4\Gamma\omega+2\zeta+\xi+|eB|-2|eB|}
-\frac{(1-\frac{1}{\chi})(\omega+i\Gamma)^2}{2\zeta+\xi+|eB|-2|eB|},
\end{array}
$$
$$
\begin{array}{cll}
z_1&=&\frac{(\omega+i\Gamma)^2
-(10\lambda_{SO}^2+2\lambda_R^2)+\zeta+\xi-|eB|}{-i4\Gamma\omega-2\zeta-2\xi+2|eB|}
-\frac{(\omega+i\Gamma)^2
-(10\lambda_{SO}^2+2\lambda_R^2)+\zeta+\xi-|eB|}{-i4\Gamma\omega-2\zeta+2|eB|}
-\frac{(\omega+i\Gamma)^2
-(10\lambda_{SO}^2+2\lambda_R^2)+\zeta+\xi-|eB|}{-2\zeta-2\xi+2|eB|}\\
&+&\frac{(\omega+i\Gamma)^2
-(10\lambda_{SO}^2+2\lambda_R^2)+\zeta+\xi-|eB|}{-2\zeta+2|eB|},
\end{array}
$$
$$
\begin{array}{cll}
z_2&=&\frac{(\omega+i\Gamma)^2
-(10\lambda_{SO}^2+2\lambda_R^2)-\zeta-\xi+|eB|}{i4\Gamma\omega-2\zeta-2\xi+2|eB|}
-\frac{(\omega+i\Gamma)^2
-(10\lambda_{SO}^2+2\lambda_R^2)-\zeta-\xi+|eB|}{i4\Gamma\omega-2\zeta+2|eB|}
-\frac{(\omega+i\Gamma)^2
-(10\lambda_{SO}^2+2\lambda_R^2)-\zeta-\xi+|eB|}{-2\zeta-2\xi+2|eB|}
\\&+&\frac{(\omega+i\Gamma)^2
-(10\lambda_{SO}^2+2\lambda_R^2)-\zeta-\xi+|eB|}{-2\zeta+2|eB|},
\end{array}
$$
$$
\begin{array}{cll}
z_3&=&\frac{(\omega+i\Gamma)^2
-(10\lambda_{SO}^2+2\lambda_R^2)-\zeta+\xi+|eB|}{i4\Gamma\omega-2\zeta+2|eB|}
-\frac{(\omega+i\Gamma)^2
-(10\lambda_{SO}^2+2\lambda_R^2)-\zeta+\xi+|eB|}{i4\Gamma\omega-2\zeta+2\xi+2|eB|}
-\frac{(\omega+i\Gamma)^2
-(10\lambda_{SO}^2+2\lambda_R^2)-\zeta+\xi+|eB|}{-2\zeta+2|eB|}
\\&+&\frac{(\omega+i\Gamma)^2
-(10\lambda_{SO}^2+2\lambda_R^2)-\zeta+\xi+|eB|}{-2\zeta+2\xi+2|eB|},
\end{array}
$$
$$
\begin{array}{cll}
z_4&=&\frac{(\omega+i\Gamma)^2
-(10\lambda_{SO}^2+2\lambda_R^2)+\zeta-\xi-|eB|}{-i4\Gamma\omega-2\zeta+2|eB|}
-\frac{(\omega+i\Gamma)^2
-(10\lambda_{SO}^2+2\lambda_R^2)+\zeta-\xi-|eB|}{-i4\Gamma\omega-2\zeta+2\xi+2|eB|}
-\frac{(\omega+i\Gamma)^2
-(10\lambda_{SO}^2+2\lambda_R^2)+\zeta-\xi-|eB|}{-2\zeta+2|eB|}
\\&+&\frac{(\omega+i\Gamma)^2
-(10\lambda_{SO}^2+2\lambda_R^2)+\zeta-\xi-|eB|}{-2\zeta+2\xi+2|eB|},
\end{array}
$$
$$
\begin{array}{cll}
z_5&=&\frac{(\omega+i\Gamma)^2
-(10\lambda_{SO}^2+2\lambda_R^2)-\zeta-\xi-|eB|}{-i4\Gamma\omega+2\zeta+2\xi+2|eB|}
-\frac{(\omega+i\Gamma)^2
-(10\lambda_{SO}^2+2\lambda_R^2)-\zeta-\xi-|eB|}{-i4\Gamma\omega+2\zeta+2|eB|}
-\frac{(\omega+i\Gamma)^2
-(10\lambda_{SO}^2+2\lambda_R^2)-\zeta-\xi-|eB|}{2\zeta+2\xi+2|eB|}
\\&+&\frac{(\omega+i\Gamma)^2
-(10\lambda_{SO}^2+2\lambda_R^2)-\zeta-\xi-|eB|}{2\zeta+2|eB|},
\end{array}
$$
$$
\begin{array}{cll}
z_6&=&\frac{(\omega+i\Gamma)^2
-(10\lambda_{SO}^2+2\lambda_R^2)+\zeta+\xi+|eB|}{i4\Gamma\omega+2\zeta+2\xi+2|eB|}
-\frac{(\omega+i\Gamma)^2
-(10\lambda_{SO}^2+2\lambda_R^2)+\zeta+\xi+|eB|}{i4\Gamma\omega+2\zeta+2|eB|}
-\frac{(\omega+i\Gamma)^2
-(10\lambda_{SO}^2+2\lambda_R^2)+\zeta+\xi+|eB|}{2\zeta+2\xi+2|eB|}
\\&+&\frac{(\omega+i\Gamma)^2
-(10\lambda_{SO}^2+2\lambda_R^2)+\zeta+\xi+|eB|}{2\zeta+2|eB|},
\end{array}
$$
$$
\begin{array}{cll}
z_7&=&\frac{(\omega+i\Gamma)^2
-(10\lambda_{SO}^2+2\lambda_R^2)+\zeta-\xi+|eB|}{i4\Gamma\omega+2\zeta+2|eB|}
-\frac{(\omega+i\Gamma)^2
-(10\lambda_{SO}^2+2\lambda_R^2)+\zeta-\xi+|eB|}{i4\Gamma\omega+2\zeta-2\xi+2|eB|}
-\frac{(\omega+i\Gamma)^2
-(10\lambda_{SO}^2+2\lambda_R^2)+\zeta-\xi+|eB|}{2\zeta+2|eB|}
\\&+&\frac{(\omega+i\Gamma)^2
-(10\lambda_{SO}^2+2\lambda_R^2)+\zeta-\xi+|eB|}{2\zeta-2\xi+2|eB|},
\end{array}
$$
$$
\begin{array}{cll}
z_8&=&\frac{(\omega+i\Gamma)^2
-(10\lambda_{SO}^2+2\lambda_R^2)-\zeta+\xi-|eB|}{-i4\Gamma\omega+2\zeta+2|eB|}
-\frac{(\omega+i\Gamma)^2
-(10\lambda_{SO}^2+2\lambda_R^2)-\zeta+\xi-|eB|}{-i4\Gamma\omega+2\zeta-2\xi+2|eB|}
-\frac{(\omega+i\Gamma)^2
-(10\lambda_{SO}^2+2\lambda_R^2)-\zeta+\xi-|eB|}{2\zeta+2|eB|}
\\&+&\frac{(\omega+i\Gamma)^2
-(10\lambda_{SO}^2+2\lambda_R^2)-\zeta+\xi-|eB|}{2\zeta-2\xi+2|eB|}.
\end{array}
$$
The Hall conductivity is then given by
\begin{equation}
\begin{array}{cll}
\sigma_{xy}=\sigma_{xy}(\Omega\rightarrow0)&=&-\frac{2e^2v_F^2|eB|sgn(eB)}{\pi^2}Im\int_{-\infty}^{\infty}
d\omega\frac{1}{4T\cosh^2\frac{\beta(\omega-\mu)}{2}}[X_H+\frac{1}{2|eB|}Y_H
+(\frac{6\lambda_{SO}\lambda_{R}}{\xi})^2Z_H]
\end{array}\label{sigxxa2}
\end{equation}
with
\begin{equation}
X_H=a_1/2-a'_1,\ \ Y_H=b_1/2-(b_1'+b_2'),\ \ Z_H=c_1/2-c_1',
\end{equation}
where
\begin{equation}
\begin{array}{cll}
a_1&=&\frac{2(1-\frac{1}{\chi})}
{(\omega+i\Gamma)^2-(10\lambda_{SO}^2+2\lambda_R^2)-\zeta+\xi+|eB|}
\frac{\omega^2+\Gamma^2}{-i4\Gamma\omega+2\zeta-\xi-|eB|+2|eB|}
+\frac{2(1+\frac{1}{\chi})}
{(\omega+i\Gamma)^2-(10\lambda_{SO}^2+2\lambda_R^2)-\zeta-\xi+|eB|}
\frac{\omega^2+\Gamma^2}{-i4\Gamma\omega+2\zeta+\xi-|eB|+2|eB|}\\
&+&\frac{\omega^2+\Gamma^2}
{(\omega+i\Gamma)^2-(10\lambda_{SO}^2+2\lambda_R^2)+\zeta}
[\frac{2(1-\frac{1}{\chi})}{-i4\Gamma\omega-2\zeta-\xi-|eB|+2|eB|}
+\frac{2(1+\frac{1}{\chi})}{-i4\Gamma\omega-2\zeta+\xi-|eB|+2|eB|}],
\end{array}
\end{equation}
\begin{equation}
\begin{array}{cll}
b_1&=&\mu_1\psi(-\frac{(\omega+i\Gamma)^2-(10\lambda_{SO}^2+2\lambda_R^2)+\zeta}{2|eB|})
-\mu_2\psi(-\frac{(\omega+i\Gamma)^2-(10\lambda_{SO}^2+2\lambda_R^2)-\zeta+\xi+|eB|}{2|eB|})
-\mu_3\psi(-\frac{(\omega+i\Gamma)^2-(10\lambda_{SO}^2+2\lambda_R^2)
-\zeta-\xi+|eB|}{2|eB|})\\
&+&\mu_4\psi(-\frac{(\omega+i\Gamma)^2-(10\lambda_{SO}^2+2\lambda_R^2)-\zeta+\xi-|eB|}{2|eB|})
+\mu_5\psi(-\frac{(\omega+i\Gamma)^2-(10\lambda_{SO}^2+2\lambda_R^2)-\zeta-\xi-|eB|}{2|eB|}),
\end{array}
\end{equation}
\begin{equation}
\begin{array}{cll}
c_1&=&\nu_1
\psi(-\frac{(\omega+i\Gamma)^2-(10\lambda_{SO}^2+2\lambda_R^2)+\zeta+\xi-|eB|}{2|eB|})
-\nu_2
\psi(-\frac{(\omega+i\Gamma)^2-(10\lambda_{SO}^2+2\lambda_R^2)+\zeta-\xi-|eB|}{2|eB|})
+\nu_3
\psi(-\frac{(\omega+i\Gamma)^2-(10\lambda_{SO}^2+2\lambda_R^2)-\zeta-\xi-|eB|}{2|eB|})\\
&-&\nu_4
\psi(-\frac{(\omega+i\Gamma)^2-(10\lambda_{SO}^2+2\lambda_R^2)-\zeta+\xi-|eB|}{2|eB|})
-\nu_5
\psi(-\frac{(\omega+i\Gamma)^2-(10\lambda_{SO}^2+2\lambda_R^2)+\zeta+\xi+|eB|}{2|eB|})
+\nu_6
\psi(-\frac{(\omega+i\Gamma)^2-(10\lambda_{SO}^2+2\lambda_R^2)+\zeta-\xi+|eB|}{2|eB|})\\
&-&\nu_7
\psi(-\frac{(\omega+i\Gamma)^2-(10\lambda_{SO}^2+2\lambda_R^2)-\zeta-\xi+|eB|}{2|eB|})
+\nu_8
\psi(-\frac{(\omega+i\Gamma)^2-(10\lambda_{SO}^2+2\lambda_R^2)-\zeta+\xi+|eB|}{2|eB|}),
\end{array}
\end{equation}
\begin{equation}
\begin{array}{cll}
a'_1&=&\alpha_0+\alpha_1\ln
[(\omega+i\Gamma)^2-(10\lambda_{SO}^2+2\lambda_R^2) -\zeta-\xi+|eB|]
+\alpha_2\ln [(\omega+i\Gamma)^2-(10\lambda_{SO}^2+2\lambda_R^2)
-\zeta+\xi+|eB|]\\&+&\alpha_3\ln
[(\omega+i\Gamma)^2-(10\lambda_{SO}^2+2\lambda_R^2) +\zeta],
\end{array}
\end{equation}
\begin{equation}
\begin{array}{cll}
b'_1&=&
\beta_1\ln\Gamma(-\frac{(\omega+i\Gamma)^2-(10\lambda_{SO}^2+2\lambda_R^2)
+\zeta}{2|eB|})+\beta_2\psi(-\frac{(\omega+i\Gamma)^2-(10\lambda_{SO}^2+2\lambda_R^2)
+\zeta}{2|eB|})
+\beta_3\phi(-\frac{(\omega+i\Gamma)^2-(10\lambda_{SO}^2+2\lambda_R^2)
+\zeta}{2|eB|}),
\end{array}
\end{equation}
\begin{equation}
\begin{array}{cll}
b'_2&=&\beta_4
\ln\Gamma(-\frac{(\omega+i\Gamma)^2-(10\lambda_{SO}^2+2\lambda_R^2)
-\zeta-\xi+|eB|}{2|eB|})+\beta_5\phi(-\frac{(\omega+i\Gamma)^2-(10\lambda_{SO}^2+2\lambda_R^2)
-\zeta-\xi+|eB|}{2|eB|})\\
&-&\beta_6\psi(-\frac{(\omega+i\Gamma)^2-(10\lambda_{SO}^2+2\lambda_R^2)
-\zeta-\xi+|eB|}{2|eB|})
+\beta_7\ln\Gamma(-\frac{(\omega+i\Gamma)^2-(10\lambda_{SO}^2+2\lambda_R^2)
-\zeta+\xi+|eB|}{2|eB|})\\
&+&\beta_8\phi(-\frac{(\omega+i\Gamma)^2-(10\lambda_{SO}^2+2\lambda_R^2)
-\zeta+\xi+|eB|}{2|eB|})-\beta_9\psi(-\frac{(\omega+i\Gamma)^2-(10\lambda_{SO}^2+2\lambda_R^2)
-\zeta+\xi+|eB|}{2|eB|})\\
&-&\beta_{10}\ln\Gamma(-\frac{(\omega+i\Gamma)^2-(10\lambda_{SO}^2+2\lambda_R^2)
-\zeta-\xi-|eB|}{2|eB|})
-\beta_{11}\phi(-\frac{(\omega+i\Gamma)^2-(10\lambda_{SO}^2+2\lambda_R^2)
-\zeta-\xi-|eB|}{2|eB|})\\&+&
\beta_{12}\psi(-\frac{(\omega+i\Gamma)^2-(10\lambda_{SO}^2+2\lambda_R^2)
-\zeta-\xi-|eB|}{2|eB|})
-\beta_{13}\ln\Gamma(-\frac{(\omega+i\Gamma)^2-(10\lambda_{SO}^2+2\lambda_R^2)
-\zeta+\xi-|eB|}{2|eB|})\\&-&
\beta_{14}\phi(-\frac{(\omega+i\Gamma)^2-(10\lambda_{SO}^2+2\lambda_R^2)
-\zeta+\xi-|eB|}{2|eB|})
+\beta_{15}\psi(-\frac{(\omega+i\Gamma)^2-(10\lambda_{SO}^2+2\lambda_R^2)
-\zeta+\xi-|eB|}{2|eB|}),
\end{array}
\end{equation}
\begin{equation}
\begin{array}{cll}
c'_1&=&
\gamma_1\ln\Gamma(-\frac{(\omega+i\Gamma)^2-(10\lambda_{SO}^2+2\lambda_R^2)+\zeta+\xi-|eB|}{2|eB|})
+\gamma_4\ln\Gamma(-\frac{(\omega+i\Gamma)^2-(10\lambda_{SO}^2+2\lambda_R^2)+\zeta-\xi-|eB|}{2|eB|})
\\&+&\gamma_7\ln\Gamma(-\frac{(\omega+i\Gamma)^2-(10\lambda_{SO}^2+2\lambda_R^2)-\zeta-\xi-|eB|}{2|eB|})
+\gamma_{10}\ln\Gamma(-\frac{(\omega+i\Gamma)^2-(10\lambda_{SO}^2+2\lambda_R^2)-\zeta+\xi-|eB|}{2|eB|})
\\&+&\gamma_{13}\ln\Gamma(-\frac{(\omega+i\Gamma)^2-(10\lambda_{SO}^2+2\lambda_R^2)+\zeta+\xi+|eB|}{2|eB|})
+\gamma_{16}\ln\Gamma(-\frac{(\omega+i\Gamma)^2-(10\lambda_{SO}^2+2\lambda_R^2)+\zeta-\xi+|eB|}{2|eB|})
\\&+&\gamma_{19}\ln\Gamma(-\frac{(\omega+i\Gamma)^2-(10\lambda_{SO}^2+2\lambda_R^2)-\zeta-\xi+|eB|}{2|eB|})
+\gamma_{22}\ln\Gamma(-\frac{(\omega+i\Gamma)^2-(10\lambda_{SO}^2+2\lambda_R^2)-\zeta+\xi+|eB|}{2|eB|})
\\&+&\gamma_2\psi(-\frac{(\omega+i\Gamma)^2-(10\lambda_{SO}^2+2\lambda_R^2)+\zeta+\xi-|eB|}{2|eB|})
+\gamma_6\psi(-\frac{(\omega+i\Gamma)^2-(10\lambda_{SO}^2+2\lambda_R^2)+\zeta-\xi-|eB|}{2|eB|})
\\&+&\gamma_9\psi(-\frac{(\omega+i\Gamma)^2-(10\lambda_{SO}^2+2\lambda_R^2)-\zeta-\xi-|eB|}{2|eB|})
+\gamma_{12}\psi(-\frac{(\omega+i\Gamma)^2-(10\lambda_{SO}^2+2\lambda_R^2)-\zeta+\xi-|eB|}{2|eB|})
\\&+&\gamma_{15}\psi(-\frac{(\omega+i\Gamma)^2-(10\lambda_{SO}^2+2\lambda_R^2)+\zeta+\xi+|eB|}{2|eB|})
+\gamma_{18}\psi(-\frac{(\omega+i\Gamma)^2-(10\lambda_{SO}^2+2\lambda_R^2)+\zeta-\xi+|eB|}{2|eB|})
\\&+&\gamma_{21}\psi(-\frac{(\omega+i\Gamma)^2-(10\lambda_{SO}^2+2\lambda_R^2)-\zeta-\xi+|eB|}{2|eB|})
+\gamma_{24}\psi(-\frac{(\omega+i\Gamma)^2-(10\lambda_{SO}^2+2\lambda_R^2)-\zeta+\xi+|eB|}{2|eB|})
\\&+&\gamma_3\phi(-\frac{(\omega+i\Gamma)^2-(10\lambda_{SO}^2+2\lambda_R^2)+\zeta+\xi-|eB|}{2|eB|})
+\gamma_5\phi(-\frac{(\omega+i\Gamma)^2-(10\lambda_{SO}^2+2\lambda_R^2)+\zeta-\xi-|eB|}{2|eB|})
\\&+&\gamma_8\phi(-\frac{(\omega+i\Gamma)^2-(10\lambda_{SO}^2+2\lambda_R^2)-\zeta-\xi-|eB|}{2|eB|})
+\gamma_{11}\phi(-\frac{(\omega+i\Gamma)^2-(10\lambda_{SO}^2+2\lambda_R^2)-\zeta+\xi-|eB|}{2|eB|})
\\&+&\gamma_{14}\phi(-\frac{(\omega+i\Gamma)^2-(10\lambda_{SO}^2+2\lambda_R^2)+\zeta+\xi+|eB|}{2|eB|})
+\gamma_{17}\phi(-\frac{(\omega+i\Gamma)^2-(10\lambda_{SO}^2+2\lambda_R^2)+\zeta-\xi+|eB|}{2|eB|})
\\&+&\gamma_{23}\phi(-\frac{(\omega+i\Gamma)^2-(10\lambda_{SO}^2+2\lambda_R^2)-\zeta+\xi+|eB|}{2|eB|})
+\gamma_{20}\phi(-\frac{(\omega+i\Gamma)^2-(10\lambda_{SO}^2+2\lambda_R^2)-\zeta-\xi+|eB|}{2|eB|})
\end{array}
\end{equation}
with $\phi(z)=\int_0^z \ln\Gamma(x)dx$, and
\begin{equation}
\begin{array}{cll}
\mu_1
&=&\frac{2(1+\frac{1}{\chi})(\omega^2+\Gamma^2)}{-i4\Gamma\omega-2\zeta-\xi+|eB|-2|eB|}
-\frac{2(1+\frac{1}{\chi})(\omega^2+\Gamma^2)}{-i4\Gamma\omega-2\zeta+\xi-|eB|+2|eB|}
+\frac{2(1-\frac{1}{\chi})(\omega^2+\Gamma^2)}{-i4\Gamma\omega-2\zeta+\xi+|eB|-2|eB|}
-\frac{2(1-\frac{1}{\chi})(\omega^2+\Gamma^2)}{-i4\Gamma\omega-2\zeta-\xi-|eB|+2|eB|},
\end{array}
\end{equation}
\begin{equation}
\begin{array}{cll}
\mu_2
=\frac{2(1-\frac{1}{\chi})(\omega^2+\Gamma^2)}{-i4\Gamma\omega+2\zeta-\xi-|eB|+2|eB|},
\ \ \mu_3
=\frac{2(1+\frac{1}{\chi})(\omega^2+\Gamma^2)}{-i4\Gamma\omega+2\zeta+\xi-|eB|+2|eB|},
\ \ \
\end{array}
\end{equation}
\begin{equation}
\begin{array}{cll}
\mu_4
=\frac{2(1+\frac{1}{\chi})(\omega^2+\Gamma^2)}{-i4\Gamma\omega+2\zeta-\xi+|eB|-2|eB|},
\ \ \mu_5
=\frac{2(1-\frac{1}{\chi})(\omega^2+\Gamma^2)}{-i4\Gamma\omega+2\zeta+\xi+|eB|-2|eB|},
\end{array}
\end{equation}
\begin{equation}
\begin{array}{cll}
\nu_1
=2\frac{(\omega+i\Gamma)^2-(10\lambda_{SO}+2\lambda_R)^2+\zeta+\xi-|eB|}
{-i4\Gamma\omega-2\zeta-2\xi+2|eB|}
-2\frac{(\omega+i\Gamma)^2-(10\lambda_{SO}+2\lambda_R)^2+\zeta+\xi-|eB|}
{-i4\Gamma\omega-2\zeta+2|eB|},
\end{array}
\end{equation}
\begin{equation}
\begin{array}{cll}
\nu_2
=2\frac{(\omega+i\Gamma)^2-(10\lambda_{SO}+2\lambda_R)^2+\zeta-\xi-|eB|}
{-i4\Gamma\omega-2\zeta+2|eB|}
-2\frac{(\omega+i\Gamma)^2-(10\lambda_{SO}+2\lambda_R)^2+\zeta-\xi-|eB|}
{-i4\Gamma\omega-2\zeta+2\xi+2|eB|},
\end{array}
\end{equation}
\begin{equation}
\begin{array}{cll}
\nu_3
=2\frac{(\omega+i\Gamma)^2-(10\lambda_{SO}+2\lambda_R)^2-\zeta-\xi-|eB|}
{-i4\Gamma\omega+2\zeta+2\xi+2|eB|}
-2\frac{(\omega+i\Gamma)^2-(10\lambda_{SO}+2\lambda_R)^2-\zeta-\xi-|eB|}
{-i4\Gamma\omega+2\zeta+2|eB|},
\end{array}
\end{equation}
\begin{equation}
\begin{array}{cll}
\nu_4
=2\frac{(\omega+i\Gamma)^2-(10\lambda_{SO}+2\lambda_R)^2-\zeta+\xi-|eB|}
{-i4\Gamma\omega+2\zeta+2|eB|}
-2\frac{(\omega+i\Gamma)^2-(10\lambda_{SO}+2\lambda_R)^2-\zeta+\xi-|eB|}
{-i4\Gamma\omega+2\zeta-2\xi+2|eB|},
\end{array}
\end{equation}
\begin{equation}
\begin{array}{cll}
\nu_5
=2\frac{(\omega+i\Gamma)^2-(10\lambda_{SO}+2\lambda_R)^2+\zeta+\xi+|eB|}
{-i4\Gamma\omega-2\zeta-2\xi-2|eB|}
-2\frac{(\omega+i\Gamma)^2-(10\lambda_{SO}+2\lambda_R)^2+\zeta+\xi-|eB|}
{-i4\Gamma\omega-2\zeta-2|eB|},
\end{array}
\end{equation}
\begin{equation}
\begin{array}{cll}
\nu_6
=2\frac{(\omega+i\Gamma)^2-(10\lambda_{SO}+2\lambda_R)^2+\zeta-\xi+|eB|}
{-i4\Gamma\omega-2\zeta-2|eB|}
-2\frac{(\omega+i\Gamma)^2-(10\lambda_{SO}+2\lambda_R)^2+\zeta-\xi+|eB|}
{-i4\Gamma\omega-2\zeta+\xi-2|eB|},
\end{array}
\end{equation}
\begin{equation}
\begin{array}{cll}
\nu_7
=2\frac{(\omega+i\Gamma)^2-(10\lambda_{SO}+2\lambda_R)^2-\zeta-\xi+|eB|}
{-i4\Gamma\omega+2\zeta+2\xi-2|eB|}
-2\frac{(\omega+i\Gamma)^2-(10\lambda_{SO}+2\lambda_R)^2-\zeta-\xi+|eB|}
{-i4\Gamma\omega+2\zeta-2|eB|},
\end{array}
\end{equation}
\begin{equation}
\begin{array}{cll}
\nu_8
=2\frac{(\omega+i\Gamma)^2-(10\lambda_{SO}+2\lambda_R)^2-\zeta+\xi+|eB|}
{-i4\Gamma\omega+2\zeta-2|eB|}
-2\frac{(\omega+i\Gamma)^2-(10\lambda_{SO}+2\lambda_R)^2-\zeta+\xi+|eB|}
{-i4\Gamma\omega+2\zeta-2\xi-2|eB|},
\end{array}
\end{equation}
$$
\begin{array}{cll}
\alpha_0&=&\frac{2(1+\frac{1}{\chi})(\omega+i\Gamma)^2}
{(2\zeta+\xi-|eB|+2|eB|)^2} +\frac{(10\lambda_{SO}^2+2\lambda_R^2)
+\zeta+\xi-|eB|}{2\zeta+\xi-|eB|+2|eB|}\frac{\frac{1}{2}(1+\frac{1}{\chi})}
{(\omega+i\Gamma)^2-(10\lambda_{SO}^2+2\lambda_R^2)-\zeta-\xi+|eB|}
+\frac{2(1-\frac{1}{\chi})(\omega+i\Gamma)^2}{(2\zeta-\xi-|eB|+2|eB|)^2}
\\&+&\frac{(10\lambda_{SO}^2+2\lambda_R^2)+\zeta-\xi-|eB|}
{2\zeta-\xi-|eB|+2|eB|}\frac{\frac{1}{2}(1-\frac{1}{\chi})}
{(\omega+i\Gamma)^2-(10\lambda_{SO}^2+2\lambda_R^2) -\zeta+\xi+|eB|}
+\frac{2(1+\frac{1}{\chi})(\omega+i\Gamma)^2}
{(2\zeta-\xi+|eB|-2|eB|)^2}\\&+&\frac{(10\lambda_{SO}^2+2\lambda_R^2)
-\zeta}{2\zeta-\xi+|eB|-2|eB|}\frac{\frac{1}{2}(1+\frac{1}{\chi})}
{(\omega+i\Gamma)^2-(10\lambda_{SO}^2+2\lambda_R^2)+\zeta}
+\frac{2(1-\frac{1}{\chi})(\omega+i\Gamma)^2}
{(2\zeta+\xi+|eB|-2|eB|)^2} +\frac{(10\lambda_{SO}^2+2\lambda_R^2)
-\zeta}{2\zeta+\xi+|eB|-2|eB|}\frac{\frac{1}{2}(1-\frac{1}{\chi})}
{(\omega+i\Gamma)^2-(10\lambda_{SO}^2+2\lambda_R^2)+\zeta},
\end{array}
$$
$$
\begin{array}{cll}
\alpha_1=\frac{2(1+\frac{1}{\chi})
[(10\lambda_{SO}^2+2\lambda_R^2)+2\zeta+3\xi-5|eB|]}{(2\zeta+\xi-|eB|+2|eB|)^2},\
\ \alpha_2=\frac{2(1-\frac{1}{\chi})
[(10\lambda_{SO}^2+2\lambda_R^2)+2\zeta-3\xi-5|eB|]}
{(2\zeta-\xi-|eB|+2|eB|)^2},
\end{array}
$$
$$
\begin{array}{cll}
\alpha_3=\frac{2(1+\frac{1}{\chi})[(10\lambda_{SO}^2+2\lambda_R^2)
-2\zeta-\xi+|eB|-2|eB|]}{(2\zeta-\xi+|eB|-2|eB|)^2}
+\frac{2(1-\frac{1}{\chi})[(10\lambda_{SO}^2+2\lambda_R^2)
-2\zeta+\xi+|eB|-2|eB|]}{(2\zeta+\xi+|eB|-2|eB|)^2},
\end{array}
$$
$$
\begin{array}{cll}
\beta_1&=&-\frac{
2(1+\frac{1}{\chi})|eB|[2(\omega+i\Gamma)^2+2\zeta+\xi-|eB|+2|eB|]}
{(2\zeta+\xi-|eB|+2|eB|)^2} -\frac{2(1-\frac{1}{\chi})
|eB|[2(\omega+i\Gamma)^2+2\zeta-\xi-|eB|+2|eB|]}
{(2\zeta-\xi-|eB|+2|eB|)^2}
\\&+&\frac{2(1-\frac{1}{\chi})|eB|[2(\omega+i\Gamma)^2+2\zeta+\xi+|eB|-2|eB|]}
{(2\zeta+\xi+|eB|-2|eB|)^2}+\frac{2(1+\frac{1}{\chi})
|eB|[2(\omega+i\Gamma)^2+2\zeta-\xi+|eB|-2|eB|]}
{(2\zeta-\xi+|eB|-2|eB|)^2},
\end{array}
$$
$$
\begin{array}{cll}
\beta_2=- \frac{(1+\frac{1}{\chi})(\omega+i\Gamma)^2} {
2\zeta+\xi-|eB|+2|eB|} -
\frac{(1-\frac{1}{\chi})(\omega+i\Gamma)^2} {
2\zeta-\xi-|eB|+2|eB|}
+\frac{(1-\frac{1}{\chi})(\omega+i\Gamma)^2} {
2\zeta+\xi+|eB|-2|eB|}
+\frac{(1+\frac{1}{\chi})(\omega+i\Gamma)^2} {
2\zeta-\xi+|eB|-2|eB|},
\end{array}
$$
$$
\begin{array}{cll}
\beta_3=-\frac{ 8|eB|^2(1+\frac{1}{\chi})}
{(2\zeta+\xi-|eB|+2|eB|)^2} - \frac{ 8|eB|^2(1-\frac{1}{\chi})} {
(2\zeta-\xi-|eB|+2|eB|)^2} +\frac{ 8|eB|^2(1-\frac{1}{\chi})} {
(2\zeta+\xi+|eB|-2|eB|)^2} +\frac{ 8|eB|^2(1+\frac{1}{\chi})} {
(2\zeta-\xi+|eB|-2|eB|)^2},
\end{array}
$$
$$
\begin{array}{cll}
\beta_4=\frac{2(1+\frac{1}{\chi})|eB|[2(\omega+i\Gamma)^2-2\zeta-\xi+|eB|-2|eB|]}
{(2\zeta+\xi-|eB|+2|eB|)^2},
\end{array}
$$
$$
\begin{array}{cll}
\beta_5=\frac{8(1+\frac{1}{\chi})|eB|^2}
{(2\zeta+\xi-|eB|+2|eB|)^2} ,\ \
\beta_6=\frac{(1+\frac{1}{\chi})(\omega+i\Gamma)^2}
{2\zeta+\xi-|eB|+2|eB|},
\end{array}
$$
$$
\begin{array}{cll}
\beta_7=\frac{2(1-\frac{1}{\chi})|eB|[2(\omega+i\Gamma)^2-2\zeta+\xi+|eB|-2|eB|]}
{(2\zeta-\xi-|eB|+2|eB|)^2},
\end{array}
$$
$$
\begin{array}{cll}
\beta_8=\frac{8(1-\frac{1}{\chi})|eB|^2}
{(2\zeta-\xi-|eB|+2|eB|)^2} ,\ \
\beta_9=\frac{(1-\frac{1}{\chi})(\omega+i\Gamma)^2}
{2\zeta-\xi-|eB|+2|eB|},
\end{array}
$$
$$
\begin{array}{cll}
\beta_{10}=\frac{2(1-\frac{1}{\chi})|eB|[2(\omega+i\Gamma)^2-2\zeta-\xi-|eB|+2|eB|]}
{(2\zeta+\xi+|eB|-2|eB|)^2},
\end{array}
$$
$$
\begin{array}{cll}
\beta_{11}=\frac{8(1-\frac{1}{\chi})|eB|^2}
{(2\zeta+\xi+|eB|-2|eB|)^2} ,\ \
\beta_{12}=\frac{(1-\frac{1}{\chi})(\omega+i\Gamma)^2}
{2\zeta+\xi+|eB|-2|eB|},
\end{array}
$$
$$
\begin{array}{cll}
\beta_{13}=\frac{2(1+\frac{1}{\chi})|eB|[2(\omega+i\Gamma)^2-2\zeta+\xi-|eB|+2|eB|]}
{(2\zeta-\xi+|eB|-2|eB|)^2},
\end{array}
$$
$$
\begin{array}{cll}
\beta_{14}=\frac{8(1+\frac{1}{\chi})|eB|^2}
{(2\zeta-\xi+|eB|-2|eB|)^2} ,\ \  \beta_{15}=
\frac{(1+\frac{1}{\chi})(\omega+i\Gamma)^2}
{2\zeta-\xi+|eB|-2|eB|},
\end{array}
$$
$$
\begin{array}{cll}
\gamma_1=\frac{4|eB|[(\omega+i\Gamma)^2-(10\lambda_{SO}^2+2\lambda_R^2)+2\zeta+\xi-2|eB|]}
{(-2\zeta+2|eB|)^2}-\frac{4|eB|[(\omega+i\Gamma)^2-(10\lambda_{SO}^2+2\lambda_R^2)+2\zeta+2\xi-2|eB|]}
{(-2\zeta-2\xi+2|eB|)^2},
\end{array}
$$
$$
\begin{array}{cll}
\gamma_2=\frac{(\omega+i\Gamma)^2-(10\lambda_{SO}^2+2\lambda_R^2)+\zeta+\xi-|eB|}
{-2\zeta-2\xi+2|eB|}-
\frac{(\omega+i\Gamma)^2-(10\lambda_{SO}^2+2\lambda_R^2)+\zeta+\xi-|eB|}
{-2\zeta+2|eB|},
\end{array}
$$
$$
\begin{array}{cll}
\gamma_3=\frac{8|eB|^2} {(-2\zeta+2|eB|)^2}-\frac{8|eB|^2}
{(-2\zeta-2\xi+2|eB|)^2},
\end{array}
$$
$$
\begin{array}{cll}
\gamma_4=\frac{4|eB|[(\omega+i\Gamma)^2-(10\lambda_{SO}^2+2\lambda_R^2)+2\zeta-\xi-2|eB|]}
{(-2\zeta+2|eB|)^2}-\frac{4|eB|
[(\omega+i\Gamma)^2-(10\lambda_{SO}^2+2\lambda_R^2)+2\zeta-2\xi-2|eB|]}
{(-2\zeta+2\xi+2|eB|)^2},
\end{array}
$$
$$
\begin{array}{cll}
\gamma_5=\frac{8|eB|^2}{(-2\zeta+2|eB|)^2}-\frac{8|eB|^2}{(-2\zeta+2\xi+2|eB|)^2},
\end{array}
$$
$$
\begin{array}{cll}
\gamma_6=\frac{(\omega+i\Gamma)^2-(10\lambda_{SO}^2+2\lambda_R^2)+\zeta-\xi-|eB|}
{-2\zeta+2\xi+2|eB|}-\frac{(\omega+i\Gamma)^2-(10\lambda_{SO}^2+2\lambda_R^2)+\zeta-\xi-|eB|}
{-2\zeta+2|eB|},
\end{array}
$$
$$
\begin{array}{cll}
\gamma_7=\frac{4|eB|[(\omega+i\Gamma)^2-(10\lambda_{SO}^2+2\lambda_R^2)-2\zeta-\xi-2|eB|]}
{(2\zeta+2|eB|)^2}-\frac{4|eB|
[(\omega+i\Gamma)^2-(10\lambda_{SO}^2+2\lambda_R^2)-2\zeta-2\xi-2|eB|]}
{(2\zeta+2\xi+2|eB|)^2},
\end{array}
$$
$$
\begin{array}{cll}
\gamma_8=\frac{8|eB|^2} {(2\zeta+2|eB|)^2}-\frac{8|eB|^2}
{(2\zeta+2\xi+2|eB|)^2},
\end{array}
$$
$$
\begin{array}{cll}
\gamma_9=\frac{(\omega+i\Gamma)^2-(10\lambda_{SO}^2+2\lambda_R^2)-\zeta-\xi-|eB|}
{2\zeta+2\xi+2|eB|}-
\frac{(\omega+i\Gamma)^2-(10\lambda_{SO}^2+2\lambda_R^2)-\zeta-\xi-|eB|}
{2\zeta+2|eB|},
\end{array}
$$
$$
\begin{array}{cll}
\gamma_{10}=\frac{4|eB|[(\omega+i\Gamma)^2-(10\lambda_{SO}^2+2\lambda_R^2)-2\zeta+\xi-2|eB|]}
{(2\zeta+2|eB|)^2}-\frac{4|eB|[(\omega+i\Gamma)^2-(10\lambda_{SO}^2+2\lambda_R^2)-2\zeta+2\xi-2|eB|]}
{(2\zeta-2\xi+2|eB|)^2},
\end{array}
$$
$$
\begin{array}{cll}
\gamma_{11}=\frac{8|eB|^2} {(2\zeta+2|eB|)^2}-\frac{8|eB|^2}
{(2\zeta-2\xi+2|eB|)^2},
\end{array}
$$
$$
\begin{array}{cll}
\gamma_{12}=\frac{(\omega+i\Gamma)^2-(10\lambda_{SO}^2+2\lambda_R^2)-\zeta+\xi-|eB|}
{2\zeta-2\xi+2|eB|}-
\frac{(\omega+i\Gamma)^2-(10\lambda_{SO}^2+2\lambda_R^2)-\zeta+\xi-|eB|}
{2\zeta+2|eB|},
\end{array}
$$
$$
\begin{array}{cll}
\gamma_{13}=\frac{4|eB|[(\omega+i\Gamma)^2-(10\lambda_{SO}^2+2\lambda_R^2)+2\zeta+2\xi+2|eB|]}
{(-2\zeta-2\xi-2|eB|)^2}-\frac{4|eB|[(\omega+i\Gamma)^2-(10\lambda_{SO}^2+2\lambda_R^2)+2\zeta+\xi+2|eB|]}
{(-2\zeta-2|eB|)^2},
\end{array}
$$
$$
\begin{array}{cll}
\gamma_{14}=\frac{8|eB|^2} {(-2\zeta-2\xi-2|eB|)^2}-\frac{8|eB|^2}
{(-2\zeta-2|eB|)^2},
\end{array}
$$
$$
\begin{array}{cll}
\gamma_{15}=\frac{(\omega+i\Gamma)^2-(10\lambda_{SO}^2+2\lambda_R^2)+\zeta+\xi+|eB|}
{-2\zeta-2|eB|}-
\frac{(\omega+i\Gamma)^2-(10\lambda_{SO}^2+2\lambda_R^2)+\zeta+\xi+|eB|}
{-2\zeta-2\xi-2|eB|},
\end{array}
$$
$$
\begin{array}{cll}
\gamma_{16}=\frac{4|eB|[(\omega+i\Gamma)^2-(10\lambda_{SO}^2+2\lambda_R^2)+2\zeta-2\xi+2|eB|]}
{(-2\zeta+2\xi-2|eB|)^2}-\frac{4|eB|[(\omega+i\Gamma)^2-(10\lambda_{SO}^2+2\lambda_R^2)+2\zeta-\xi+2|eB|]}
{(-2\zeta-2|eB|)^2},
\end{array}
$$
$$
\begin{array}{cll}
\gamma_{17}=\frac{8|eB|^2} {(-2\zeta+2\xi-2|eB|)^2}-\frac{8|eB|^2}
{(-2\zeta-2|eB|)^2},
\end{array}
$$
$$
\begin{array}{cll}
\gamma_{18}=\frac{(\omega+i\Gamma)^2-(10\lambda_{SO}^2+2\lambda_R^2)+\zeta-\xi+|eB|}
{-2\zeta-2|eB|}-
\frac{(\omega+i\Gamma)^2-(10\lambda_{SO}^2+2\lambda_R^2)+\zeta-\xi+|eB|}
{-2\zeta+2\xi-2|eB|},
\end{array}
$$
$$
\begin{array}{cll}
\gamma_{19}=\frac{4|eB|[(\omega+i\Gamma)^2-(10\lambda_{SO}^2+2\lambda_R^2)-2\zeta-2\xi+2|eB|]}
{(2\zeta+2\xi-2|eB|)^2}-\frac{2|eB|[(\omega+i\Gamma)^2-(10\lambda_{SO}^2+2\lambda_R^2)-2\zeta-\xi+2|eB|]}
{(2\zeta-2|eB|)^2},
\end{array}
$$
$$
\begin{array}{cll}
\gamma_{20}=\frac{8|eB|^2} {(2\zeta+2\xi-2|eB|)^2}-\frac{8|eB|^2}
{(2\zeta-2|eB|)^2},
\end{array}
$$
$$
\begin{array}{cll}
\gamma_{21}=\frac{(\omega+i\Gamma)^2-(10\lambda_{SO}^2+2\lambda_R^2)-\zeta-\xi+|eB|}
{2\zeta-2|eB|}-
\frac{(\omega+i\Gamma)^2-(10\lambda_{SO}^2+2\lambda_R^2)+\zeta-\xi+|eB|}
{2\zeta+2\xi-2|eB|},
\end{array}
$$
$$
\begin{array}{cll}
\gamma_{22}=\frac{4|eB|[(\omega+i\Gamma)^2-(10\lambda_{SO}^2+2\lambda_R^2)-2\zeta+2\xi+2|eB|]}
{(2\zeta-2\xi-2|eB|)^2}-\frac{4|eB|[(\omega+i\Gamma)^2-(10\lambda_{SO}^2+2\lambda_R^2)-2\zeta+\xi+2|eB|]}
{(2\zeta-2|eB|)^2},
\end{array}
$$
$$
\begin{array}{cll}
\gamma_{23}=\frac{8|eB|^2} {(2\zeta-2\xi-2|eB|)^2}-\frac{8|eB|^2}
{(2\zeta-2|eB|)^2},
\end{array}
$$
$$
\begin{array}{cll}
\gamma_{24}=\frac{(\omega+i\Gamma)^2-(10\lambda_{SO}^2+2\lambda_R^2)-\zeta+\xi+|eB|}
{2\zeta-2|eB|}-
\frac{(\omega+i\Gamma)^2-(10\lambda_{SO}^2+2\lambda_R^2)-\zeta+\xi+|eB|}
{2\zeta-2\xi-2|eB|}.
\end{array}
$$
The Eqs. (\ref{sigxxa1}) and (\ref{sigxxa2}) are further rewritten
as Eqs. (\ref{sigxx}) and (\ref{sigxy}).
\end{widetext}

\end{document}